\documentclass[aps,prd,superscriptaddress,twocolumn,amsfonts,floatfix,showpacs]{revtex4}

\usepackage{graphicx}
\usepackage{amsmath}
\def\prl{Phys. Rev. Lett.}
\def\prd{Phys. Rev. D}
\def\cqg{Class. Quantum Grav.}

\def\CovDev{D}
\def\Res{{\mathcal R}}

\begin{document}
   
\title{An alternative approach to solving the Hamiltonian constraint}

\author{Thomas W. Baumgarte}
\altaffiliation{Also at Department of Physics, University of Illinois at Urbana-Champaign, 
Urbana, IL 61801}
\affiliation{Department of Physics and Astronomy, Bowdoin College,
  Brunswick, ME 04011}

\begin{abstract}
Solving Einstein's constraint equations for the construction of black hole initial data requires handling the black hole singularity.  Typically, this is done either with the excision method, in which the black hole interior is excised from the numerical grid, or with the puncture method, in which the singular part of the conformal factor is expressed in terms of an analytical background solution, and the Hamiltonian constraint is then solved for a correction to the background solution that, usually, is assumed to be regular everywhere.  We discuss an alternative approach in which the Hamiltonian constraint is solved for an inverse power of the conformal factor.  This new function remains finite everywhere, so that this approach  requires neither excision nor a split into background and correction.  In particular, this method can be used without modification even when the correction to the conformal factor is singular itself.  We demonstrate this feature for rotating black holes in the trumpet topology.
\end{abstract}

\pacs{04.20.Ex, 04.25.D-, 04.25.dg, 04.70.Bw}

\maketitle

\section{Introduction}

Constructing initial data in general relativity requires solving the constraint equations of Einstein's field equations (see, e.g.,~\cite{Alc08,BauS10}).   Under the assumption of conformal flatness and maximal slicing, solutions to the {\em momentum constraint} describing boosted or spinning back holes can be expressed analytically in terms of Bowen-York solutions \cite{BowY80}.  These solutions can then be inserted into the {\em Hamiltonian constraint}, which, in general, has to be solved numerically for the conformal factor.

Any numerical method employed to solve the Hamiltonian constraint for black hole data has to accommodate the presence of black hole singularities.   One approach is the {\em excision method}, in which the black hole interior is excised from the numerical grid, and suitable boundary conditions are imposed on the black hole horizon (see, e.g., \cite{Coo91,Coo94,CooP04}).  An alternative is the {\em puncture method}, in which the singular parts of the solution are expressed in terms of an analytical background solution, and the Hamiltonian constraint is solved for a regular correction to the background solution \cite{BeiO94,BraB97}.  

Here we discuss an alternative approach that requires neither excision nor a decomposition into background and correction (even though the latter is probably desirable in terms of numerical accuracy).  Specifically, we consider solving the Hamiltonian constraint for an inverse power of the conformal factor.   This approach, which is similar to an approach that has become extremely successful in solving Einstein's evolution equations (e.g.~\cite{CamLMZ06}), appears to be promising in the context of Einstein's constraint equations as well.  The new function representing the conformal factor remains finite everywhere and can be solved for directly.  We present numerical examples and compare with both analytical and independent numerical results.  An important advantage of this approach over the puncture method is that it can be used without modification even when, in the puncture method, the correction diverges as fast as the background solution itself.  We demonstrate this feature for rotating black holes in the trumpet topology \cite{ImmB09,HanHO09}.   We expect that this property may be important for applications that relax the assumption of conformal flatness, since a non-vanishing deviation from conformal flatness may lead, in the context of the puncture method, to singular corrections to the analytic background terms.   

\section{Basic equations}  
\label{sec:BE}

In vacuum, and under the assumption of maximal slicing and conformal flatness, the Hamiltonian constraint reduces to
\begin{equation} \label{Ham1}
\bar \CovDev^2 \psi = - \frac{1}{8} \psi^{-7} \bar A_{ij} \bar A^{ij},
\end{equation}
where $\psi$ is the conformal factor and $\bar A^{ij} = \psi^{10} A^{ij}$ is the conformally rescaled, trace-free part of the extrinsic curvature.  Also, $\bar \CovDev_i$ is the covariant derivative operator with respect to the conformally related metric $\bar \gamma_{ij} = \psi^4 \gamma_{ij}$, where $\gamma_{ij}$ is the physical spatial metric.  Under the assumption of conformal flatness, $\bar \gamma_{ij} = \eta_{ij}$, where $\eta_{ij}$ is the flat metric any coordinate system, the operator $\bar \CovDev^2 \equiv \bar \gamma^{ij} \bar \CovDev_i \bar \CovDev_j$ reduces to the flat Laplace operator.

Now consider a new function
\begin{equation} \label{Omega}
\Omega \equiv \psi^{-n},
\end{equation}
where $n$ is a constant that we will later choose to be a positive integer.  Since the conformal factor  $\psi$ typically diverges at the black hole singularity, this choice makes $\Omega$ go to zero and remain finite there.  The Laplace operator acting on $\Omega$ then satisfies
\begin{equation}
\bar \CovDev^2 \Omega = \left(1 + \frac{1}{n} \right) \Omega^{-1} \bar \gamma^{ij} 
	\bar \CovDev_i \Omega \bar \CovDev_j \Omega
	- n \Omega^{1 + 1/n} \bar \CovDev^2 \psi.
\end{equation}
Inserting (\ref{Ham1}) we can now express the Hamiltonian constraint in terms of $\Omega$,
\begin{equation} \label{Ham2}
\bar \CovDev^2 \Omega = \left(1 + \frac{1}{n} \right) 
	\frac{\bar \CovDev^i \Omega \bar \CovDev_i \Omega}{\Omega}
	+ \frac{n}{8}\, \Omega^{1 + 8/n} \bar A_{ij} \bar A^{ij}.
\end{equation}
Evidently we recover the Hamiltonian constraint in its original form (\ref{Ham1}) for $n = -1$.

Different choices for the power $n$ can be considered.  One possibility would be to choose $n$ in such a way that  that the first term on the right-hand side of (\ref{Ham2}) remains finite at the black hole singularity, where $\Omega$ vanishes.  Assuming that $\psi$ diverges with $r^{-m}$ at the singularity, where $r$ is the isotropic radius, $\Omega$ scales with $r^{nm}$, and the first term on the right-hand side of (\ref{Ham2}) with $r^{nm - 2}$.  This term remains finite as $r \rightarrow 0$ if $n \geq 2/m$.   For the minimum value, $n = 2/m$, we always have $\Omega \propto r^2$ close to the singularity.  

The scaling of $\psi$ close to the singularity depends on the slicing of the black hole.  For so-called wormhole data, for example, $\psi$ diverges with $r^{-1}$, so that $m = 1$.  Here we will focus on trumpet data, for which $\psi \propto r^{-1/2}$ close to the singularity (see \cite{HanHPBO06,HanHOBGS06,BauN07,HanHOBO08} and equation (\ref{psi0}) below).  The above argument would then suggests that we should choose $n \geq 4$.   Despite these considerations, $n=4$ may not be the best choice.  As we will discuss in more detail below, we found better results for $n=2$, even though some terms in (\ref{Ham2}) diverge with $1/r$ for this choice.


Before proceeding we linearize the Hamiltonian constraint (\ref{Ham2}) as follows.  Denoting the approximate solution after $l$ iteration steps with $\Omega^l$, we search for a correction $\delta \Omega \ll \Omega^l$ so that $\Omega^{l+1} = \Omega^l + \delta \Omega$ solves the equation.  
Denoting the residual of equation (\ref{Ham2}) for $\Omega^l$ with $\Res^l$,
\begin{equation}
\Res^l \equiv \bar \CovDev^2 \Omega^l  
	- \left(1 + \frac{1}{n} \right) 
	\frac{\bar \CovDev^i \Omega^l \bar \CovDev_i \Omega^l}{\Omega^l} - 
	\frac{n}{8} (\Omega^l)^{1 + 8/n} \bar A^2, 
\end{equation}
where we have abbreviated $\bar A^2 = \bar A_{ij} \bar A^{ij}$,
the linearized Hamiltonian constraint (\ref{Ham2}) becomes
\begin{eqnarray} \label{Hamlin2}
& & \bar \CovDev^2 (\delta \Omega) -
\left(1 + \frac{1}{n} \right) \frac{2}{\Omega^l} 
\bar \gamma^{ij} \bar \CovDev_i \Omega^l \bar \CovDev_j \delta \Omega  +  \nonumber \\
& & 
\left\{ \left(1 + \frac{1}{n} \right) 
\frac{\bar \CovDev^i \Omega^l \bar \CovDev_i \Omega^l}{(\Omega^l)^2}  - 
\left(\frac{n}{8} + 1 \right) (\Omega^l)^{8/n} \bar A^2 \right\} \delta \Omega \nonumber \\
& & ~~~~~~~ = - \Res^l.
\end{eqnarray}
Equation (\ref{Hamlin2}) is a linear equation that can be solved iteratively for $\delta \Omega$ until the norm of the residual $\Res^l$ has dropped below a desired tolerance.

\section{Numerical examples}

\subsection{Schwarzschild}

\begin{figure}[t]
\includegraphics[width=3in]{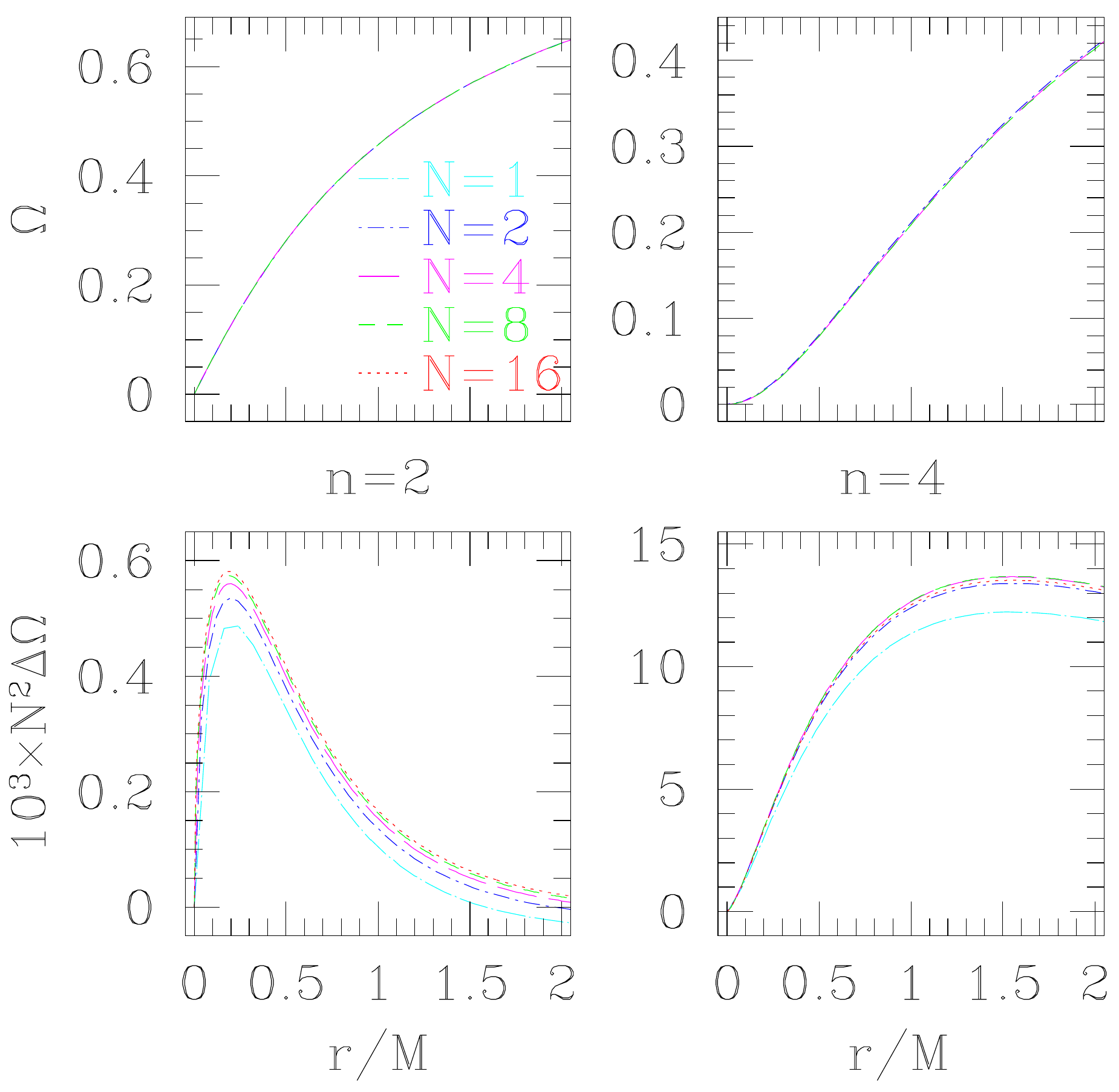}
\caption{Numerical solutions $\Omega$ for a Schwarzschild black hole, for $n=2$ and $n=4$.  Here we imposed the outer boundary at $R_{\rm out} = 16 M$ and used $100N + 1$ gridpoints.  The upper panels shows the solution for different values of $N$.  As expected, $\Omega$ scales with $r^{n/2}$ at the center.  The lower panel shows the rescaled errors $N^2 \Delta \Omega$, demonstrating second-order convergence to the analytical solution, even in the neighborhood of the singularity.}
\label{Fig1}
\end{figure}

As a first test of this scheme we solve the Hamiltonian constraint (\ref{Ham2}) in spherical symmetry to construct the Schwarzschild solution.   

We adopt a finite-difference method and use a uniform vertex-centered grid so that the first grid point is at $r = 0$.   We set $\Omega = 0$ at $r = 0$, and set $\Omega$ to its analytical value $\psi_0^{-n}$ at the outer boundary $R_{\rm out}$ of the grid.  Finite-differencing the operator on the left-hand side of equation (\ref{Hamlin2}) results in a tridiagonal matrix that can be solved with elementary methods.  

In order to construct maximally sliced trumpet-data, we adopt 
\begin{equation} \label{A_0}
\bar A^{ij}_0 = \frac{3 \sqrt{3} M^2}{4 r^3} \left( \bar \gamma^{ij} - 3 n^i n^j \right),
\end{equation}
where $n^i = x^i/r$ is the spatial normal vector pointing away from the center of the black hole at $r = 0$, and $M$ is the total mass-energy of the black hole \footnote{Without splitting $\Omega$ into a background and a correction term, the mass of the resulting black hole is specified by the extrinsic curvature.  Wormhole data, for which the extrinsic curvature vanishes, therefore do require a split of $\Omega$ into background and correction, since otherwise the mass remains undetermined.}. The analytical solution for $\psi_0$  can also be given analytically, albeit only in parametric form (see \cite{BauN07}).  In the neighborhood of the singularity, $\psi_0$ is given by
\begin{equation} \label{psi0}
\psi_0 = \left( \frac{3 M}{2r} \right)^{1/2}
\end{equation}
to leading order in $r$.  Throughout this paper, $M$, denotes the mass of the background Schwarzschild solution.

In Figure \ref{Fig1} we show numerical results as a function of the radius $r$ for two choices $n=2$ and $n=4$.  The upper panel shows the solutions $\Omega$ for different grid resolutions; as the resolution increases, the numerical solutions approach the analytical solution $\Omega_0 = \psi_0^{-n}$.  As expected, $\Omega$ scales with $r^{n/2}$ close to the center.  The lower panel shows the errors $\Delta \Omega \equiv \Omega - \Omega_0$, rescaled with the square of the grid-spacing.  These results demonstrate that the scheme is second-order accurate even in the vicinity of the black hole singularity.  They also demonstrate that this method can be adopted without decomposing the solution into a background and a correction term, even though we will use such a decomposition in the next sections.

The choice $n=4$ has the appealing features that the first term on the right hand side of (\ref{Ham2}) remains finite, as we discussed above, and that $\Omega$ is smooth at the center.  However, Fig.~\ref{Fig1} shows that for $n=2$ the numerical errors are smaller.  We have also found that for $n=4$ our iterative scheme failed to converge for rapidly spinning black holes (see Section \ref{sec:rotBH} below).  We will therefore focus on $n=2$ for the remainder of this paper.

 \subsection{Boosted black holes}

\begin{figure}[t]
\includegraphics[width=3in]{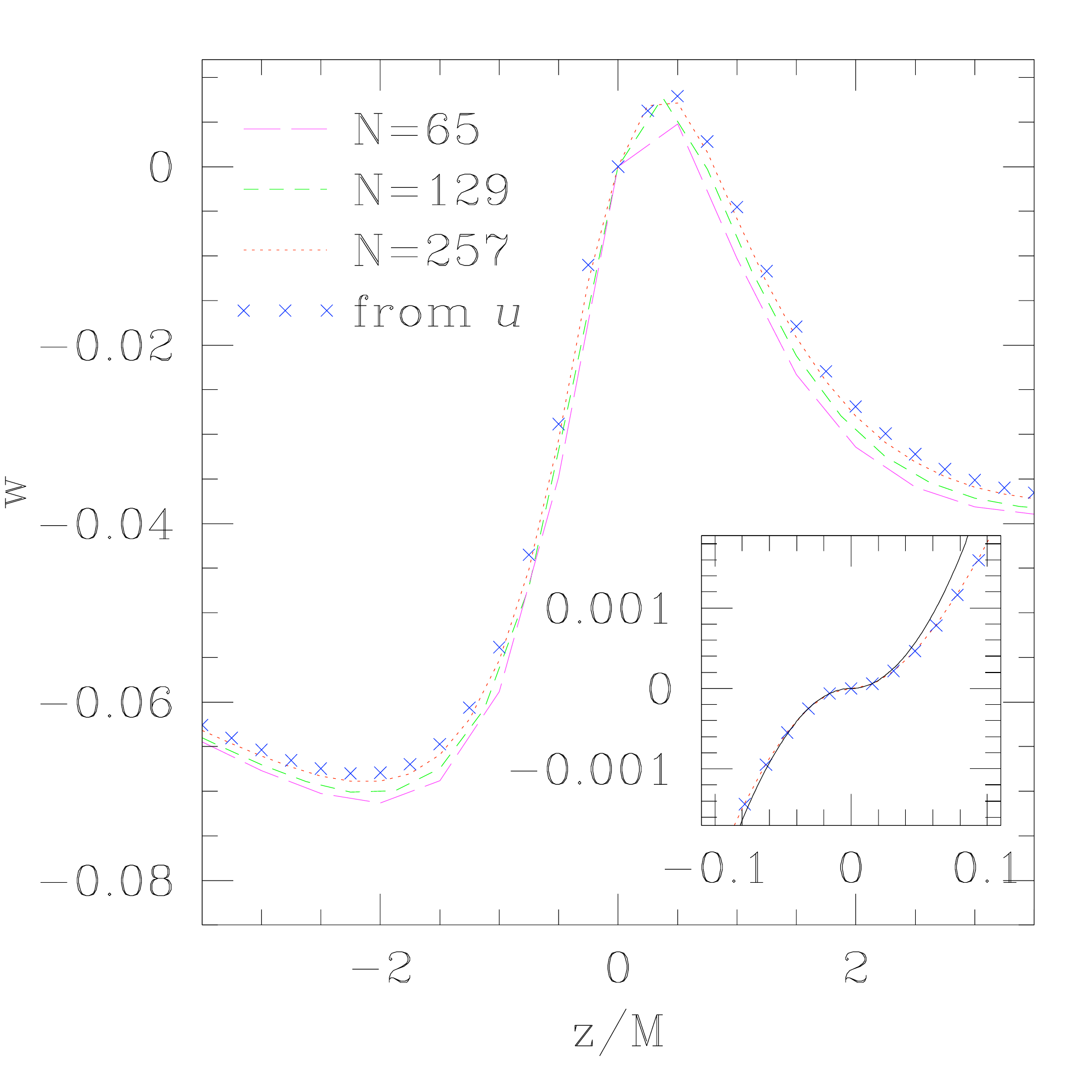}
\caption{Numerical solutions $w$ for boosted black holes with momentum $P^z/M = 1.0$.  The lines in the large plot show results for $w$, obtained for $n=2$, on three different numerical grids with increasing grid resolution and more distant outer boundaries.  For $N=65$, the outer boundaries were imposed at $X_{\rm out} = 16M$, for $N = 129$ at $X_{\rm out} = 24M$, and for $N =257$ at $X_{\rm out} = 32M$.  Also included, as crosses, are the results as computed from the Hamiltonian constraint in its original from (\ref{Ham1}) with $N = 257$.   The insert shows higher-resolution results (obtained with $N = 257$ and $X_{\rm out} = 2 M$) of the region around the black hole center.  This graph also includes the leading-order analytical result (\ref{w_boost}) as a solid line.}
\label{Fig2}
\end{figure}

We construct boosted trumpet black holes by adding to the Schwarzschild extrinsic curvature (\ref{A_0}) a Bowen-York solution representing a black hole with momentum $P^i$,
\begin{equation} \label{A_BY_P}
\bar A^{ij}_P = \frac{3}{2 r^2} \left( P^i n^j + P^j n^i - (\bar \gamma^{ij} - n^i n^j) n_k P^k \right)
\end{equation}
(since the momentum constraint is linear, the sum of two solutions is still a solution).   In the interest of numerical accuracy, we decompose $\Omega$ as 
\begin{equation} \label{Omega_decomp}
\Omega = \Omega_0 + w,
\end{equation} 
where $\Omega_0$ is the analytical Schwarzschild solution, and $w$ a correction.  This way the largest terms in the Hamiltonian constraint can be computed from the analytical solution, and only the correction $w$ needs to be treated with finite-differencing.  The iteration for $w$ still works in the same way as before if we replace $\Omega^l$ in (\ref{Hamlin2}) with $\Omega_0 + w^l$, and $\delta \Omega$ with $\delta w$.

We now solve (\ref{Ham2}), using the iteration (\ref{Hamlin2}), on a three-dimensional, uniform cartesian grid of $N^3$ gridpoints, with the help of both Cactus and PETSc software.   We again adopt a vertex-centered grid, fix $w = 0$ at the center, and impose a $1/r$ fall-off condition at the outer boundaries on a square $x_{\rm out} = y_{\rm out} = z_{\rm out} = \pm X_{\rm out}$.

Boosted trumpet black holes have previously been constructed with the puncture method by solving the Hamiltonian constraint in its original form (\ref{Ham1}) (see \cite{ImmB09,HanHO09}, see also \cite{BucPB09} for a calculation using the excision method).  In that approach, the conformal factor $\psi$ is decomposed as $\psi = \psi_0 + u$.  Given a solution $u$, we can compute the corresponding $w$ from (\ref{Omega_decomp}),
\begin{equation} \label{w_from_u}
w = \Omega - \Omega_0 = (\psi_0 + u)^{-n} - \psi_0^{-n}.
\end{equation}
For boosted black holes, we can therefore compare the results from the new method discussed here with independent numerical results.  Moreover, as shown in \cite{ImmB09}, regular solutions for $u$ for boosted trumpet black holes in the neighborhood of the puncture are given, to leading order in $r$, by
\begin{equation} \label{u_boost}
u_P = - \frac{1}{3 \sqrt{2}} \frac{P}{M} \left( \frac{r}{M} \right)^{1/2} \cos \theta,
\end{equation}
where $\cos \theta = n_i \hat P^i$.  Inserting this, together with (\ref{psi0}), into (\ref{w_from_u}) we find
\begin{equation} \label{w_boost}
w_P = n \frac{2^{n/2}}{3^{(n+3)/2}} \frac{P}{M} \left(\frac{r}{M} \right)^{n/2 + 1} \cos \theta
\end{equation} 
to lowest order in $r$.  

In Fig.~\ref{Fig2} we show numerical results for a black hole boosted with a momentum $P^z/M = 1.0$.  The graph shows that, as both the resolution and the distance to the outer boundaries are increased, the results for $w$ approach those computed with the puncture method from $u$.   Note that $w$ and $u$ feature different asymptotic behavior as $r \rightarrow \infty$.  By imposing a $1/r$ fall-off on both functions at a finite $X_{\rm out}$, we suppress different higher-order terms.  Therefore, the results from the two approaches only approach each other as both the numerical resolution and the location of the outer boundaries are increased.  The inset in Fig.~\ref{Fig2} shows higher-resolution results for the region around the singularity, and demonstrates that the numerical results, for both $w$ and $u$, approach the analytical result (\ref{w_boost}) as $r \rightarrow 0$ \footnote{The analytical solution in the neighborhood for $r = 0$ arises from the particular solution of the Hamiltonian constraint (see \cite{ImmB09}), which is independent of the outer boundaries.  A comparison of the solutions in the immediate neighborhood of $r=0$ is therefore not affected by the location of the outer boundaries.}.

\subsection{Spinning black holes}
\label{sec:rotBH}

\begin{figure}[t]
\includegraphics[width=2.5in]{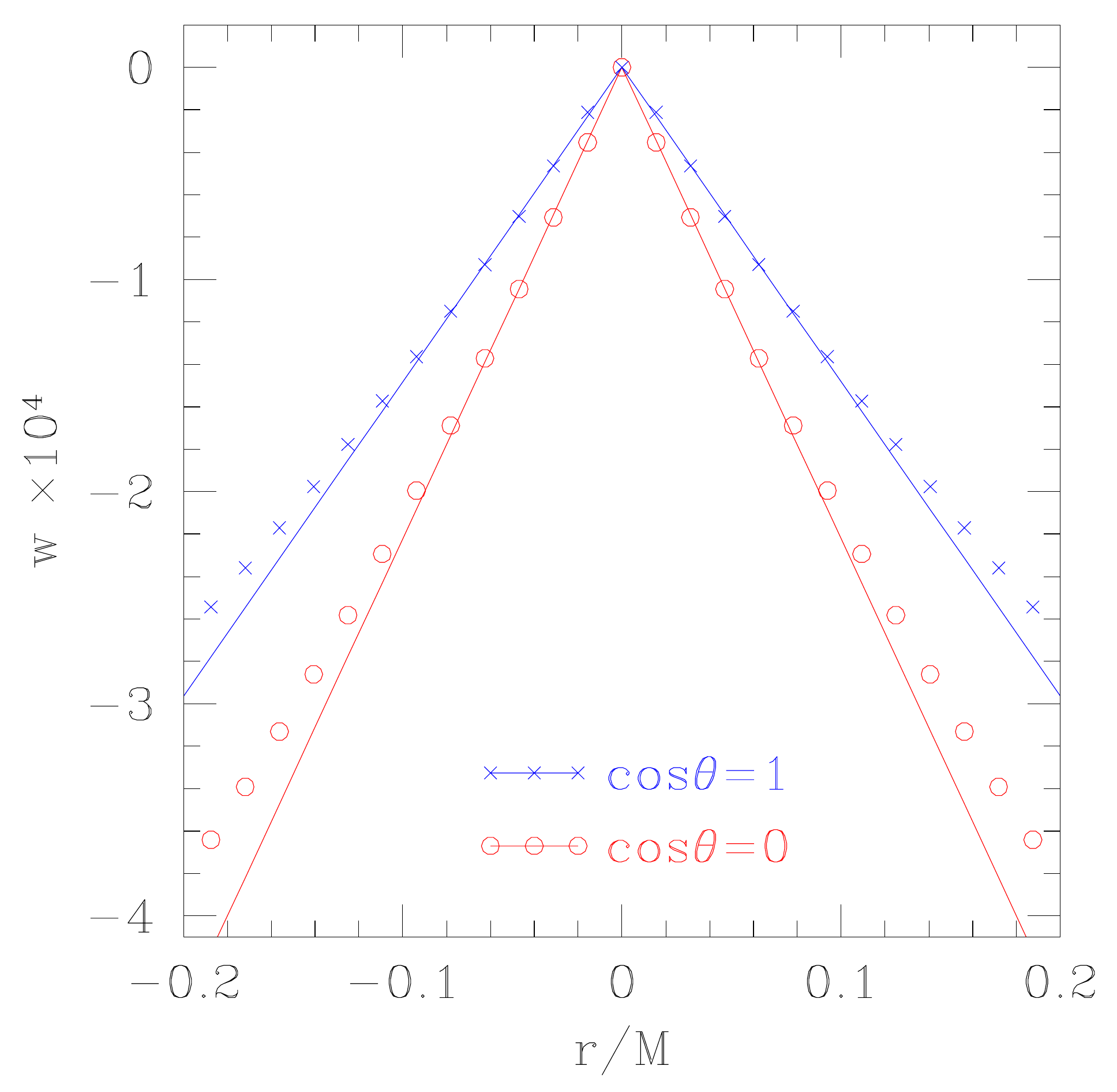}
\caption{Solutions $w$ for a black hole spinning with angular momentum $J^z/M^2 = 0.1$, using $n=2$.  We show numerical results (dots) in the vicinity of the singularity, together with the leading-order analytical result (\ref{w_spin}) (solid lines), both along the direction of the spin ($\cos \theta = 1$) and a direction orthogonal to the spin ($\cos \theta = 0$).  The numerical results were obtained with $N=257$ and $X_{\rm out} = 2M$.}
\label{Fig3}
\end{figure}

\begin{figure}[t]
\includegraphics[width=3in]{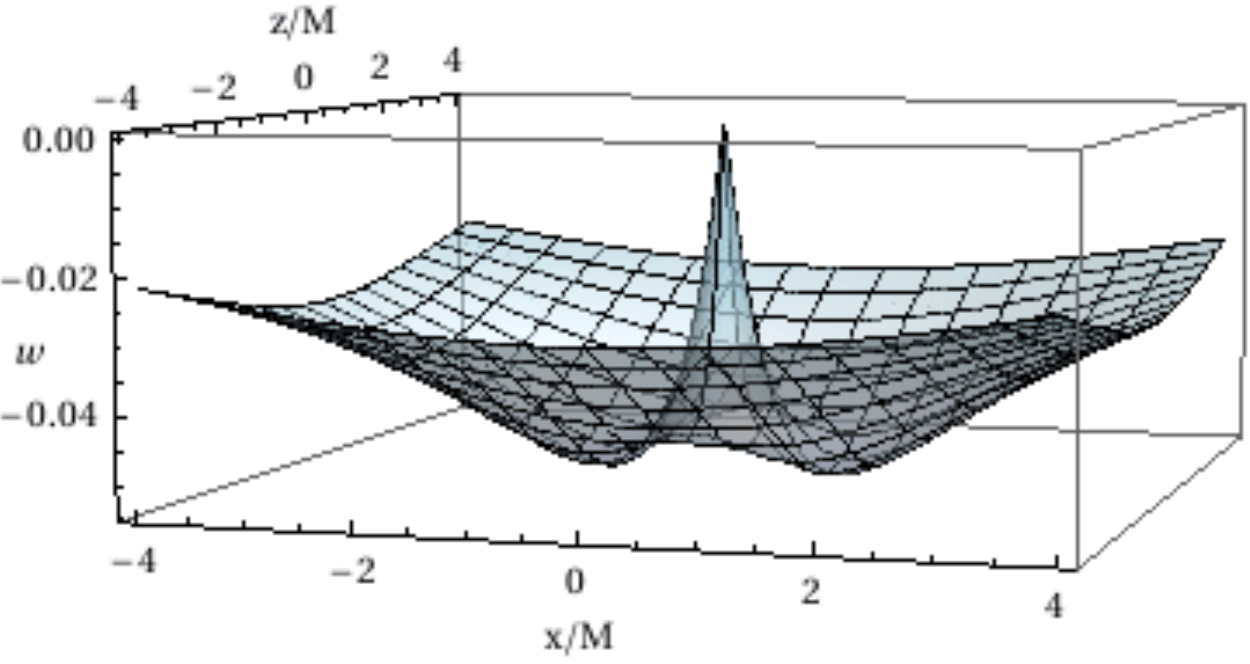}
\caption{Numerical results for $w$ for a black hole spinning with angular momentum $J^z/M^2 = 1.0$ in the $x$-$z$ plane, obtained with $N = 129$, $X_{\rm out} = 4M$ and $n=2$.}
\label{Fig4}
\end{figure}

Spinning black holes can be constructed by adding to the Schwarzschild extrinsic curvature (\ref{A_0}) a Bowen-York solution representing a black hole with spin angular momentum $J^i$,
\begin{equation} \label{A_BY_S}
\bar A^{ij}_S = \frac{6}{r^3} n^{(i} \bar \epsilon^{j)kl} J_k n_l.
\end{equation}
As a consequence of $\bar A^{ij}_S$ diverging with $r^{-3}$ (rather than $r^{-2}$ for the boosted solutions (\ref{A_BY_P})), corrections $u$ to $\psi_0$ now diverge with $r^{-1/2}$, i.e.~they diverge with the same power of $r$ as the background $\psi_0$ itself.  In the neighborhood of the singularity, corrections $u$, to leading order in both $r/M$ and $J/M^2$, are given by
\begin{equation} \label{u_spin}
u_S = \frac{1}{12} \left( \frac{J}{M^2} \right)^2 \left( \frac{2 M}{3 r} \right)^{1/2} \left(3 - \cos^2 \theta \right),
\end{equation}
where $\cos \theta = n_i \hat J^i$ (see \cite{ImmB09}).

Given that the corrections $u$ are not regular, the puncture method in its original form breaks down.  The equations can still be solved if the singular behavior of $u$ is scaled out, but this leads to rather complicated expressions (see \cite{HanHO09}).   The method proposed here, however, can still be used without any modification (other than using $\bar A^{ij}_S$ instead of $\bar A^{ij}_P$).   Inserting (\ref{u_spin}) and (\ref{psi0}) into (\ref{w_from_u}) we now have
\begin{equation} \label{w_spin}
w_S = - n \frac{2^{n/2-1}}{3^{n/2+2}} \left( \frac{J}{M^2} \right)^2 \left( \frac{r}{M} \right)^{n/2} 
\left(3 - \cos^2 \theta \right).
\end{equation}
to leading order in both $r/M$ and $J/M^2$.  Not surprisingly, this solution scales with $r^{n/2}$, just like the background term $\Omega_0$. 

In Fig.~\ref{Fig3} we show high-resolution results, for $n=2$, in the vicinity of the center for a black hole with angular momentum $J^z/M^2 = 0.1$.  For this sufficiently small value of $J$, the numerical results (crosses and circles) approach the center as predicted by the analytical result (solid lines, see eq.~(\ref{w_spin})).  Note that $w$ remains finite everywhere, even if, for $n=2$, derivatives are discontinuous at $r=0$.  However, since we can set the solution to zero there, and never need to evaluate any derivatives at the center, this does not affect the numerical scheme.  In Fig.~\ref{Fig4} we also show a surface graph of $w$ for a larger value of the angular momentum ($J^z/M^2 = 1.0$).  

\section{Discussion}

We discuss an approach to solving the Hamiltonian constraint that requires neither excision nor a decomposition into background and correction terms for the construction of black hole initial data.  Specifically, we solve the Hamiltonian constraint for an inverse power of the conformal factor.   The resulting function then remains regular everywhere, and vanishes at the location of the black hole singularities.  We present numerical examples and compare with both analytical and independent numerical results.  An important advantage of this method is that it can handle cases for which, in the puncture method, corrections diverge at the singularity as fast as the background term itself.  We demonstrate this feature for spinning trumpet black holes, and expect that this property may be important for applications in which the assumption of conformal flatness is relaxed \footnote{But see \cite{HanHBGS07} for an example in which the traditional puncture method can be used even in the absence of conformal flatness.}.

We also experiment with different powers $n$ in the rescaling $\Omega = \psi^{-n}$.  As we discuss in Section \ref{sec:BE}, the choice $n = 4$ leads to some appealing properties of the equation and the solutions; we nevertheless found $n=2$ more suitable for our iteration scheme.  

In the puncture method, corrections $u$ that diverge at the singularity, if they exist, are suppressed by the assumption of regularity there \footnote{Unless the method is modified to allow for singular corrections, as for rotating trumpet black holes, see \cite{HanHO09}.}.   In the approach discussed here, such solutions would not be suppressed automatically, since they also satisfy $w = 0$ at the center.  We therefore suspect that a second branch of solutions, corresponding to singular corrections $u$, might exist.   Analyzing the uniqueness of solutions, and the properties of any other branches of solutions, might make an interesting subject for future investigations.

\acknowledgments
 
It is a pleasure to thank Ken Dennison for his help with Cactus software, and Mark Hannam for his comments on an earlier draft of this paper.  This work was supported in part by NSF Grant PHY-1063240 to Bowdoin College.

\end{document}